\documentclass[fleqn,usenatbib]{mnras}

\usepackage{graphicx}
\usepackage{bm}
\usepackage{amsmath}
\usepackage{multirow}
\usepackage{multicol}
\usepackage{amssymb}
\usepackage{color}

\def\psidm{$\psi$DM}
\def\Msun{M_{\odot}}
\def\Lsun{L_{\odot}}
\def\cvir{c_{\rm vir}}
\def\Mvir{M_{\rm vir}}
\def\rvir{r_{\rm vir}}
\def\rs{r_{\rm s}}
\def\rhos{\rho_{\rm s}}
\def\rhocr{\rho_{\rm cr}}
\def\mB{m_{\rm B}}
\def\Mhalo{M_{\rm halo}}
\def\Msol{M_{\rm sol}}
\def\rsol{r_{\rm sol}}

\def\rcore{r_{\rm c}}
\def\Mstars{M_{\rm stars}}
\def\fstars{f_{\rm stars}}
\def\Lstars{L_{\rm stars}}
\def\RHL{R_{\rm HL}}
\def\tH{t_{\rm H}}

\newcommand{\bd}{\begin{displaymath}}
\newcommand{\ed}{\end{displaymath}}
\newcommand{\be}{\begin{equation}}
\newcommand{\ee}{\end{equation}}
\newcommand{\beaa}{\begin{eqnarray*}}
\newcommand{\eeaa}{\end{eqnarray*}}
\newcommand{\bea}{\begin{eqnarray}}
\newcommand{\eea}{\end{eqnarray}}

\newcommand{\sref}[1]{Section~\ref{#1}}
\newcommand{\aref}[1]{Appendix~\ref{#1}}
\newcommand{\fref}[1]{Figure~\ref{#1}}
\newcommand{\fsref}[1]{Figures~\ref{#1}}
\newcommand{\tref}[1]{Table~\ref{#1}}
\newcommand{\eref}[1]{Equation~(\ref{#1})}
\newcommand{\esref}[1]{Equations~(\ref{#1})}

\title[\psidm\ with stars]{How do stars affect \psidm\ halos?}
\author[James H.H. Chan et al.]{James H.H. Chan$^{1}$$^{,}$$^{2}$$^{,}$$^{3}$, Hsi-Yu Schive$^{4}$, Tak-Pong Woo$^{1}$$^{,}$$^{5}$, Tzihong Chiueh$^{1}$$^{,}$$^{6}$$^{,}$$^{7}$
\thanks{E-mail:chiuehth@phys.ntu.edu.tw}\\
$^{1}$Department of Physics, National Taiwan University, 10617, Taipei, Taiwan\\
$^{2}$Max-Planck-Institut f{\"u}r Astrophysik, Karl-Schwarzschild-Str. 1, 85748 Garching, Germany\\
$^{3}$Laboratoire d'Astrophysique, Ecole Polytechnique F{\'e}d{\'e}rale de Lausanne (EPFL), Observatoire de Sauverny, CH-1290 Versoix, Switzerland\\
$^{4}$National Center for Supercomputing Applications, Urbana, IL, 61801, USA\\
$^{5}$National Applied Research Laboratories, 10636, Taipei, Taiwan\\
$^{6}$Institute of Astrophysics, National Taiwan University. 10617, Taipei, Taiwan\\
$^{7}$Center for Theoretical Physics, National Taiwan University, 10617, Taipei, Taiwan\\
}

\begin{document}

\date{}

\pagerange{\pageref{firstpage}--\pageref{lastpage}} \pubyear{2018}

\maketitle

\label{firstpage}

\begin{abstract}
\label{abstract}
Wave dark matter (\psidm) predicts a compact soliton core and a granular halo in every galaxy.  
This work presents the first simulation study of an elliptical galaxy by including both stars and \psidm, focusing on the systematic changes of the central soliton and halo granules.  
With the addition of stars in the inner halo, we find the soliton core consistently becomes more prominent by absorbing mass from the host halo than that without stars, and the halo granules become ``non-isothermal", ``hotter" in the inner halo and ``cooler" in the outer halo, as opposed to the isothermal halo in pure \psidm\ cosmological simulations.  
Moreover, the composite (star+\psidm) mass density is found to follow a $r^{-2}$ isothermal profile near the half-light radius in most cases.  
Most striking is the velocity dispersion of halo stars that increases rapidly toward the galactic center by a factor of at least $2$ inside the half-light radius caused by the deepened soliton gravitational potential, a result that compares favorably with observations of elliptical galaxies and bulges in spiral galaxies. 
However in some rare situations we find a phase segregation turning a compact distribution of stars into two distinct populations with high and very low velocity dispersions; while the high-velocity component mostly resides in the halo, the very low-velocity component is bound to the interior of the soliton core, resembling stars in faint dwarf spheroidal galaxies. 
\end{abstract}


\section{Introduction}
\label{sec:intro}
Solitons of wave dark matter, \psidm, are ground states of Bose-Einstein condensates (BEC) under self-gravity.
(See \cite{Marsh16} and \cite{HuiEtal17} for comprehensive reviews of \psidm.)
After dissipationless violent relaxation of a finite mass, only a small fraction of total mass is condensed into the ground state and the majority is in excited states distributed over a much extended volume as a halo, as found in e.g. \citet*[][SCB14 hereafter]{SchiveEtal14} and \citet{LinEtal18}.

In such a dissipationless establishment of quasi-equilibrium, a soliton is centrally located surrounded by a massive halo, and the soliton's size $\rcore$ is related to the halo velocity dispersion $\sigma$ via $\rcore \sigma \sim \hbar/\mB$, where $\mB$ is the bosonic particle mass.
We call this relation the nonlocal uncertainty principle \citep{SchiveEtal14b} since the soliton and the halo are not co-spatial.
Nonlocal uncertainty principle has a simple physical meaning.
If one may assign an effective temperature to the soliton, with the local uncertainty principle to determine the soliton effective thermal speed, the nonlocal uncertainty principle is nothing but equilibration of temperature between the soliton and the halo.

Discovery of this nonlocal uncertainly principle is important, in that it allows one to predict the soliton mass for a given halo mass \citep{SchiveEtal14b,SchwabeEtal16}.
As an example, the soliton mass is about $10^9\Msun$ located in the central $150$~pc of the Milky Way for \psidm\ particle mass of $10^{-22}$~eV, and it may explain the possible missing mass within the central $100$~pc \citep{PortailEtal17}.
However, when we take into account abundant baryons in the inner halo (within a few kpc), it makes one wonder how severe these additional baryons may affect the soliton and the inner halo.
To formulate the question more precisely to the present study, considering that when dissipative gases fall into the gravitational potential and stay in the inner halo in forming stars, \textit{will the original relation between the soliton and the halo remains the same, and if not, to what extent it will change?}
This is one question to be addressed in this work.

The answer is not intuitively obvious.
One may argue that the gravity of additional baryons increases the velocity dispersion in the inner that constantly interacts with the soliton, and the soliton should have a size still determined by the velocity dispersion of a inner halo through the nonlocal uncertainty principle.
Alternatively, one may also argue that the soliton can exist isolation without any halo and hence the nonlocal uncertainty principle may not necessarily apply, given that dissipation has been involved in bringing in additional mass where the system energy must be lower in favor of the lowest energy ground state.

We will try to answer this question via a series of controlled simulations, and a few unexpected new features in addition to the question posed above are found.  In this work, we address the simpliest case, the elliptical galaxy, and approximate baryons by stars.
The simulation solves individual dynamical equations of \psidm\ and of collisionless stars coupled by the common gravity.   
We initialize a distribution of cold stars (to be detailed in \sref{sec:issues}) in the inner halo of a well-formed dark matter halo.
Cold stars immediately rush into the core region followed by random crossing at the center.
After a small fraction of Hubble time, stars acquire velocity dispersion and gradually settle into a dynamical quasi-equilibrium along with the halo.
These warm stars are still distributed only in the inner halo.
We then look into how the soliton and the halo velocity dispersion may have changed in the final state.

It has been known that the velocity dispersion in the \psidm\ halo manifests itself in the form of short-lived granules with a size, given by the local uncertainty principle \citep{SchiveEtal14b}.
Hence the real-space representation of halo velocity dispersion $\sigma$ is the inverse granule size.
In our quest for the answer, a comparison of sizes of the central soliton and granules provides a quantitative measure to address the aforementioned question.
Given that, a second question may arise.  
\textit{Will the granules change their sizes non-uniformly across the inner halo where stars are concentrated?}
(As a reference, in cosmological simulations with pure \psidm, granules are of roughly the same size throughout the inner halo.)
This question is relevant for answering the first question as we need to know what to compare if the target of comparison becomes nonuniform.

A third question of considerable astrophysical interest is \textit{whether or not stars and dark matter can be well coupled when fully relaxed.}
In a \psidm\ halo, the dynamical coupling between these two species can possibly be stronger than that in a conventional cold dark matter (CDM) halo, as a \psidm\ halo is filled with fluctuating density granules.
When so, the composite mass of the two components ought to approach an isothermal profile even though each individual component does not.
This issue is motivated by recent observational studies \citep{KoopmansEtal09,BarnabeEtal11,SonnenfeldEtal13}, employing strong lensing and star kinematics, that reveal total mass densities to follow isothermal profiles in inner galaxies.
Though these are massive galaxies, we see no reason why this tendency does not preserve in less massive galaxies as long as stellar mass is a non-negligible fraction of the combined mass in the inner halo.
This issue has previously been examined with CDM simulations, but only to find the combined mass density follows a power law with a power index $-(2.4-2.5)$ in purely dynamical simulations \citep{DuttonEtal15}.
Only after additional gas and star formation physics, such as stellar feedback, is incorporated can the power index turn to approach $-2$ \citep{DuttonEtal15,XuEtal17}.

The final question is again from the perspective of observations.  
\textit{Can we identify any signature of the soliton from the stellar kinematics?
If so, how does it compare with observations?} 

This work is organized in the following way.
In \sref{sec:simulation}, we describe the setup of our simulation.
Some fundamental dynamical issues of a \psidm\ halo and a soliton are addressed in \sref{sec:issues}.
The stellar kinematics and 2D projection are discussed in \sref{sec:stellar_kinematics} in connection with observations.
We conclude in \sref{sec:conclusion}.
In \aref{app:scaling} we show how to scale low mass halos to $\sim 10$ times more massive halos, both in the small galaxy range, with and without stars.
In \aref{app:slope}, we demonstrate more composite halos having isothermal mass profiles from the simulations.
In this work, we choose $\Omega_{\rm m} = 0.284$, $\Omega_{\Lambda} = 1-\Omega_{\rm m}$, $h = 0.7$, and the boson mass $\mB=8.0\times10^{-23} {\rm eV}/c^2$.

\section{Simulation}
\label{sec:simulation}

The \psidm\ halos are extracted from cosmological simulations.
The halo then evolves together with an initially cold stellar distribution located in the inner halo.
This initial stellar condition to some extent mimics a centrally located cold gas which totally converts itself into stars in a star burst of an elliptical galaxy.
The simulation consists of two parts:
1) solving \psidm\ wave function in a static universe using the spectral method with a uniform mesh \citep{Woo&Chiueh09,MoczEtal17}, and 2) solving trajectories of stars as particles following the Newtonian dynamics.

Throughout this work, we conduct simulations mostly with $512^3$-pixel spatial resolution with $2.5\times10^7$ particles as stars.
Only in some special cases when we suspect the spatial resolution may be marginal do we check results with  $1024^3$-pixel resolution simulations.

\subsection{\psidm}
\label{subsec:psidm}

The evolution of the \psidm\ model is governed by the Schr{\"o}dinger-Poisson (SP) equation \citep{Seidel&Suen90,Widrow&Kaiser93}:
\be
i\hbar\frac{\partial \psi}{\partial t} = -\frac{\hbar^2}{2\mB}\nabla^2\psi+\mB \Phi\psi
\ee
and
\be
\nabla^2\Phi = 4\pi G \rho,
\ee
where $\psi$ is the wave function, $\Phi$ the gravitational potential,
$\mB$ the mass of dark matter boson, and the mass density $\rho= \mB |\psi|^2$.
In this work, we ignore the Hubble expansion and hence consider a static universe in the vicinity of a bound halo.
This set of equations is invariant to a similarity transformation $r, t, \rho \to \lambda r, \lambda^2 t, \lambda^{-4}\rho$.
It implies a scaling relation that the soliton mass $\Msol$ is inversely proportional to the soliton size $\rsol$, or
\be
\Msol\cdot\rsol = {\rm constant}
\label{eqn:mr}
\ee
where ${\rm constant}\propto \mB^{-2}$ \citep{SchiveEtal14b}.  
In the next section, we will use $\Msol\cdot\rsol$ as an indicator to check whether the central mass bump is a soliton.  
Here, $\Msol \equiv M(<\rsol)$ is the enclosed mass within the radius $\rsol$ at which the density is $1/20$ of peak density $\rho_{\rm peak}$. 
Note that in this work we introduce a new definition for the soliton width $\rsol$ so as to capture most of the soliton mass as opposed to the soliton core width $\rcore$ of SCB14.
We find $\rsol\sim 2.2\rcore$ and $\Msol$ is about 90\% of the total soliton mass. 
More features of soliton are described in \sref{subsec:soliton_stars}.

To form a halo from the initial linear power spectrum, a uniform mesh resolution is inadequate to resolve both compact cores and aggregation of several halos, as illustrated in \cite{Woo&Chiueh09}.
This difficulty has been tackled to some extent by a highly optimized AMR framework: GAMER \citep{SchiveEtal10,SchiveEtal17}, featuring an efficient solution to integrating the Schr{\"o}dinger-Poisson (SP) equation with graphic processing units (GPUs) for computation acceleration.
Even with these optimizations, GAMER can only carry out cosmology simulations of a limited volume ($2$~Mpc$^3$ in the comoving frame), within a reasonable run time (few weeks) and computing resources ($32$ units of GPU), producing the biggest halo about $10^{11}\Msun$.

To perform follow-up simulations, we extract only two dwarf-galaxy scale halos, of mass $5.0\times10^{9}\Msun$ and $1.7\times10^{10}\Msun$, respectively, from GAMER cosmology simulations.
We primarily use the less massive halo for the majority of tests.
We do so for the following reasons.

First, one should not take the quoted halo mass literally, since the simulation in this work has no background density and we can interpret the halo mass as we please.
This arbitrariness of interpretation is however constrained by a ratio: the soliton peak height to the halo peak height: ``the shape parameter", which is determined by cosmological simulations and generally more massive halos have larger shape parameters.  
But we find empirically that in some range of halo mass, the shape parameter is rather constant. 
For a $5.0\times10^{9}\Msun$ halo, the shape parameter is about $100$, but for a $7.0\times 10^{10}\Msun$ halo obtained by the same structure formation simulation (the most massive halo in Figure 3 of SCB14) the shape parameter is $150$.  
Two shape parameters are quite similar and therefore the two inner halos are similar.
Furthermore, we place initial stars of a given mass fraction and distribution in a inner halo in reference to the soliton throughout this work.
As a result, the soliton, the inner halo and the stars can be self-similar for halos of different masses.
Thus the galaxy under investigation should be interpreted as having mass in between the approximate range of the above two quoted values, which we call the inner halo scaling relation.  
This halo scaling without stars is demonstrated by explicitly comparing $5$ halos extracted from the aforementioned cosmological simulations in the scaling range, and with stars demonstrated by examining two halos, $5.0\times10^9\Msun$ and $1.7\times10^{10}\Msun$, as an example.
The detailed discussion of inner halo scaling relation is presented in \aref{app:scaling}.

Second, this study aims to conduct many different simulation runs, and for this reason we must have quick turnovers.
To achieve it, our strategy, unlike GAMER, implements all calculations of \psidm\, including the Poisson solver, in one single GPU with uniform meshes.
Without additional complicated data structures of AMR, one can port a uniform-mesh code to GPU relatively straightforwardly and achieve a significant improvement of performance.
Taking the $512^3$-pixel computational domain as an example, the time span of advancing one time step is as short as $<1.3$~sec for \psidm\ simulations, and typically we need $10^4$ time steps.
Simultaneously we perform the trajectory calculation of stars with the multi-core CPU so as to share the computing load.
However, such uniform-grid simulations have a couple of trade-offs.
It limits the viable halo mass range to $10^{10}\Msun$ or so in order to resolve the compact core.
(See \cite{SchiveEtal14b} for the halo mass-core mass relation.)
It also prevents one from looking into situations such as halo mergers since the needed volume is prohibitively large.

Third, our primary interest is to explore the wave features, such as soliton and halo granules, those are absent in CDM.   
From the soliton-halo scaling and the non-local uncertainty principle, these wave features are more prominent in low-mass halos than in high-mass halos.   
Therefore, the focus of this work is placed on the small to dwarf galaxies.   
For example, a typical galaxy cluster has granules of size about $10$~pc and these granules are so small that they are indistinguishable from CDM discrete particles from the simulation perspective.

The wave function is advanced by a unitary transformation $\psi(t+\Delta t)=e^{-i\Phi\Delta t} e^{-iK\Delta t} \psi(t)$, where $K(=-\hbar^2\nabla^2/2\mB)$ and $\Phi$ are kinetic energy and gravitational potential energy operators, respectively acting on spectral and coordinate spaces.
Gravitational potential and kinetic energy operators are calculated via CUDA Fast Fourier Transform (CUFFT)\footnote{https://developer.nvidia.com/cufft} with the periodic boundary condition.
The simulation box is $87.5$~kpc/$h$ on each side, which is large enough to minimize the artifact arising from the periodic boundary condition.
We choose the pixel size $0.17$~kpc/$h$ for $512^3$ runs, which enables simulation runs to resolve the innermost soliton.
The temporal resolution is also chosen to be fine enough to resolve the matter-wave oscillation of every wavelength.

\subsection{Stars}
\label{subsec:stars}

The stellar initial condition is quite model-dependent ultimately related to primordial gas evolution.
It is conceivable that baryonic gases are promptly brought into the inner halo and condensed into stars as a result of losing pressure support by radiation cooling.
During the stellar evolution, a fraction of stellar mass is lost to stellar winds.
A full account of gas and stellar dynamics is well beyond the scope of work.
We thus ignore how gases might affect a halo during the stage when they were brought into the inner halo, and consider only after the epoch when stars are formed.
We assume all stars are created cold in a star burst in the inner halo, and we also assume gases are a negligible component of the baryonic content in the inner halo.   
We also do not consider the angular momentum in the stars, thus excluding the spiral galaxies.  
The initial stellar distribution is few times wider than, or comparable to, the central soliton, and 
detailed stellar setups are to be presented in \sref{subsec:soliton_stars} and \sref{subsec:5.0e09}.

The particle code adopts the standard particle-in-cell method in a uniform mesh and advance time with the Euler method.
Although the integration is only first order accurate, we demonstrate that the total energy remains conserved with at most a few percent error in all simulations.
Time steps are chosen such that none of the particles can travel more than one cell size in one time step to ensure the numerical accuracy.
Note that \psidm\ and particles share the same time steps.

The star particle number is fixed to $2.5\times10^7$ in all runs.
Half of these stars are confined within a radius of 10-25 pixels from the center in $512^3$ runs and hence it has more than sufficient mass resolution to ensure collisionless particle dynamics.

\subsection{CDM}
\label{subsec:cdm}

To compare the difference from the \psidm\ model, we conduct the CDM simulations with the same stellar distribution. 
The CDM halos are constructed following the NFW profile:
\be
\rho_{\rm NFW}(r)= {\rhos \over (r/\rs)(1+r/\rs)^2},
\ee
where $\rhos$ is the characteristic density and $\rs$ is the scaled radius,
which is defined by
\be
\rs = {\rvir \over \cvir},
\ee
with $\rvir$ being the virial radius and $\cvir$ the concentration parameter.
We define the relation between the total mass (virial mass) $\Mvir=\Mhalo$ and $\rvir$ as
\be
\Mvir = {4\pi\over 3} \rvir^3 \rhocr \Delta_{\rm od},
\ee
where $\rhocr$ is the present critical density and $\Delta_{\rm od}$ the overdensity.
The characteristic density, $\rhos$, can be obtained from the integrated mass within virial radius by following
\be
\Mvir = 4 \pi \rhos \rs^3 \left[\ln(1+\cvir)-{\cvir\over1+\cvir}\right].
\ee
The halos follow the $M$-$c$ relation from \cite{Dutton&Maccio14}:
\be
\log\cvir = a+b\log\left({\Mvir/[10^{12} h^{-1}M_\odot}]\right).
\ee
We choose $a=1.025$, $b=-0.097$, and $\Delta_{\rm od}=104.2$ for this work.
The velocity for each CDM particle is given by the Gaussian random number with one standard deviation as the velocity dispersion $\sigma(r) = \sqrt{\sigma_r(r)^2+\sigma_t(r)^2}$, where $\sigma_r(r)$ and $\sigma_t(r)$ are radial and tangential velocity dispersions respectively \citep{Lokas&Mamon01}.
The radial velocity dispersion can be derived from the Jeans equation
\be
\frac{1}{\rho}\frac{\rm d}{{\rm d} r}(\rho \sigma_r^2)+2\beta\frac{\sigma_r^2}{r}=-\frac{{\rm d}\Phi}{{\rm d}r},
\ee
where the anisotropy of the velocity dispersions is described as 
\be
\beta = 1- \frac{\sigma_t(r)^2}{2\sigma_r(r)^2}.
\ee
In this work, we investigate the inner region of dark matter haloes which is isotropic \citep{DiemandEtal04}, and hence we choose $\beta=0$.
The simulation code for CDM is the same as that for stars which adopts the particle-in-cell method.
The number of CDM particles is $512^3$ for all cases.

\section{Fundamental of baryon$+$\psidm\ halo: Time Irreversibility}
\label{sec:issues}
We address the first and second questions listed in \sref{sec:intro} with several tests.
The less massive halo ($5.0\times10^9\Msun$) is in use for these tests.

\subsection{Single soliton$+$stars}
\label{subsec:soliton_stars}
\begin{figure}
\includegraphics[width=0.35\textwidth, angle=-90]{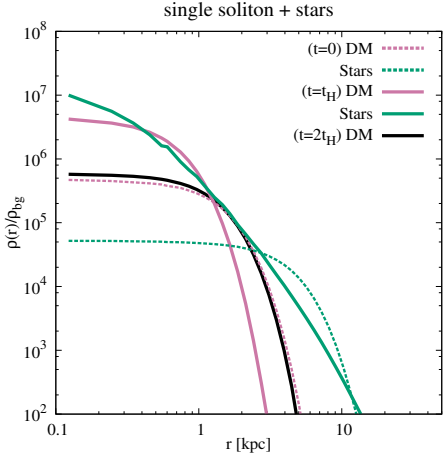}
\caption{
The density profile of single soliton with stars.
The initial density distribution of soliton/stars is shown in the red/green dot curve.
The total mass of soliton is $3.3\times10^8\Msun$ and the total mass of stars is $10^9\Msun$.
The initial cold stellar distribution is $3$ times wider than the soliton.
The density distribution of soliton/stars after evolving in one Hubble time $\tH$ is shown in the red/green solid curve.  
At $\tH$ the soliton becomes narrower by a factor of $0.5$ with the peak amplitude increased by an order of magnitude.
We then decrease the stellar mass gradually to zero from $\tH$ to $1.3\tH$, and keep relaxing the soliton until $2\tH$.
The soliton returns to the initial state as drawn in the black solid curve, showing time reversibility of the soliton.
}
\label{fig:soliton_stars}
\end{figure}

We first test the simplest case, a single soliton with stars to understand the action of stars on the soliton.
The initial profile of an isolated soliton is given by SCB14:
\be
\rho_{\rm sol}(r) = {\rho_{\rm peak} \over \left[1+9.1\times10^{-2}(r/\rcore)^2\right]^8 } \Msun{\rm pc}^{-3}.
\label{eqn:soliton}
\ee
as shown by the red dot line in \fref{fig:soliton_stars}.
Here, $\rcore$ is defined as the solitonic core radius when the soliton density drops by a half from the peak value.
In this formula, the peak density and the solitonic core radius obey a scaling relation:
\be
\rho_{\rm peak}= 1.9 (\mB/10^{-23} {\rm eV})^{-2}(\rcore/{\rm kpc})^{-4}
\label{eqn:soliton_scaling}
\ee
for an isolated soliton as well as for solitons surrounded by \psidm\ halos.
This scaling relation also implies that the product of soliton mass and radius remains constant when the mass and radius may individually change as shown in \eref{eqn:mr}.

In this test, we choose the total mass of soliton as $3.3\times10^8\Msun$ ($\Msol=3.0\times10^8\Msun$).
The initial cold stellar distribution is chosen to be $3$ times wider than the soliton and we scale the stellar particle mass to achieve the total mass of stars $9.0\times10^8\Msun$, as shown in the green dot line of \fref{fig:soliton_stars}.
The profile of the soliton after evolving with stars in a Hubble time $\tH$ is shown by the red solid line.
We then gradually turn off the stellar mass from $\tH$ to $1.3\tH$ to test the time reversibility. 
This adiabatic process is neccessary to prevent from the violent change of the soliton's state. 
When the stellar mass becomes zero we keep relaxing the soliton until $2\tH$.

One observes that the initially compressed soliton returns to the initial state after stars are removed, as shown in the black solid line, showing that time reversibility is respected.
At $\tH$, we have a more compact relaxed stellar distribution than the initial, exerting additional gravity to the soliton.
The soliton becomes narrower by a factor of $0.5$ with the peak amplitude increased by an order of magnitude, a configuration not strictly satisfying the soliton scaling \eref{eqn:mr} and \esref{eqn:soliton_scaling}.
This is expected since the soliton scaling is valid only with no external gravity, and having the star gravity the scaling can no longer hold.

This test illustrates two points.
First, the soliton can respond to the environmental gravity by adjusting itself size and its shape, but it can no longer be described by the universal soliton profile given in \eref{eqn:soliton}.
Second, as the soliton is a coherent object, it can time-reversibly respond to environmental changes.

\subsection{Soliton-halo$+$stars}
\label{subsec:5.0e09}

Here, we consider two halos of $5.0\times 10^9\Msun$ and $1.7\times 10^{10}\Msun$ for our tests, with an understanding of the halo scaling that these halos can be of any mass in between approximately few times $10^9\Msun$ and $10^{11}\Msun$. 
The initial cold star distributions are set up to be: (Case~A) $3$ times wider than the soliton, (Case~B) $6$ times wider than the soliton, and (Case~C) half of the soliton.  
The halo masses are $5.0\times 10^9\Msun$ for Cases~A and C, and $1.7\times10^{10}\Msun$ for Case~B.
The mass ratios of stars relative to the soliton are also different: $\fstars\equiv\Mstars/\Msol=3$ for Case~A, $\fstars=6$ for Case~B, and $\fstars=1.5$ for Case~C, where $\Mstars$ is the total stellar mass.

We choose the initial stellar configuration in this range so as to bring the relaxed configurations close to observations.
Take the halo mass $\Mhalo=7\times 10^{10}\Msun$ as an example.
It has a bare soliton core radius $\rcore\approx 0.4$~kpc and the soliton mass $\Msol=10^9\Msun$ according to the soliton-halo relation.
Observations show such a galaxy has the half-light radius $\RHL\simeq1$~kpc on average \citep{vanDokkumEtal09}, and thus $\RHL/\rcore\simeq 3$, and the stellar luminosity $\Lsun\sim 10^9\Lsun$ that roughly obeys the relation between $\Mhalo$ and $\Lstars$ \citep{Vale&Ostriker04}:
\be
\Lstars=5.7\times10^9 \frac{M_{11}^4}{\left(0.57+M_{11}^{0.86}\right)^{4.35}} \Lsun/h^2
\label{eqn:ltomh}
\ee
where $M_{11}\equiv\Mhalo/(10^{11}\Msun/h)$. 
Assuming the stellar mass-to-light ratio $3 \Msun/\Lsun$ \citep{Oguri06}, we have a stellar mass $\Mstars = 3.0\times10^9\Msun$, and hence $\fstars = 3$.

Now we go back to the halo with $5.0\times 10^9\Msun$ of Caes~A, which has $\fstars=3$ and a bare soliton core radius $\rcore\approx 1$kpc. 
It turns out that the relaxed halo of Case~A has $\RHL\approx 3.4$~kpc and $\RHL/\rcore\simeq 3.3$, and thus consistent with the $7.0\times 10^{10}\Msun$ galaxy.
For less (more) massive halos, the ratio of $\fstars$ are smaller (greater).  Hence
Case~C corresponds to a smaller halo of $5.0\times 10^{10}\Msun$ and Case~B to a more massive halo of $1.2\times 10^{11}\Msun$.  
It should be noted that $\RHL$ is roughly $1$~kpc for elliptical galaxies in this mass range with large scatters \citep{Graham&Guzman03,Dabringhausen&Kroupa13}.

\subsubsection{Case~A-1}

We shall examine Case~A extensively.
The simulation proceeds with immediate turn-on of stars at $t=0$.
The \psidm\ halo is interacting with stars till one Hubble time $\tH$.
(Here the fiducial halo mass is $M_{\rm h0}=5.0\times10^{9}\Msun$, and for higher mass halos, the time is shorter rescaled by the factor $(\rho_{\rm solX}/\rho_{\rm sol0})^{-1/2}=(M_{\rm hX}/M_{\rm h0})^{-2/3}$.)
After then, we decrease the stellar mass gradually to zero from $\tH$ to $1.3\tH$, and let the halo keep relaxing until $2\tH$.
The case is to test the reversibility of the soliton in the presence of halo.

\begin{figure*}
\includegraphics[width=1.0\textwidth, angle=0]{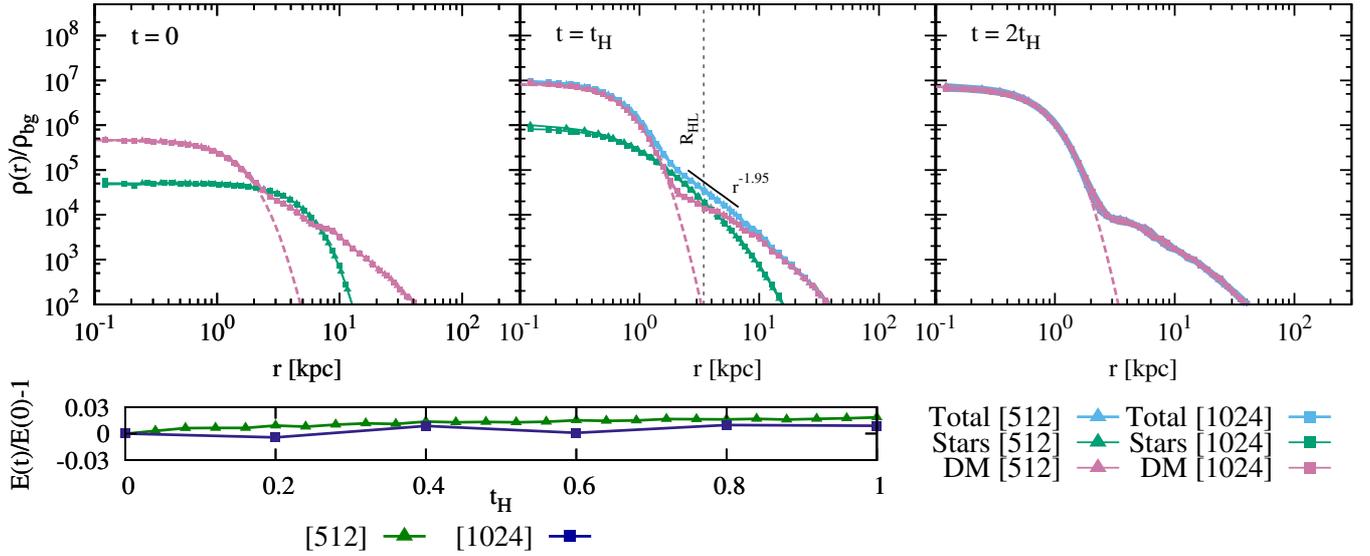}
\caption{
Density profile of Case~A-1.
The initial stellar distribution is $3$ times wider than the central soliton and has a star-to-soliton mass ratio $\fstars=3$ as shown in the top-left panel.
The halo mass is $5.0\times10^{9}\Msun$.
The \psidm\ halo and stars are relaxed at $\tH$ as shown in the top-middle panel. 
We decrease the stellar mass gradually to zero from $\tH$ to $1.3\tH$, and then we keep relaxing the \psidm\ halo until $2\tH$ as shown in the top-right panel.
The triangle and square curves are from the $512^3$ and $1024^3$ simulations respectively.
DM/Stars/Total profile is plotted in a red/green/cyan curve.
As soon as the run starts, the peak of solitonic core quickly increases about an order of magnitude.
Even when stars become massless at $2\tH$, the soliton remains in the same configuration as that at $\tH$.
The analytical solutions of solitons are given in dot curves, following \eref{eqn:soliton} with two inter-dependent parameters: $\rho_{\rm peak}$ and $\rcore$.
The bottom panel shows the time evolution of $E(t)/E(0)-1$, where $E(t)$ is the total energy at any time and $E(0)$ is the initial energy of either $512^3$ (dark green) or $1024^3$ (dark blue) resolution.
The half-light radius $\RHL$ is the 2D projected radius enclosing half of total stellar mass.
}
\label{fig:5.0e09x3}
\end{figure*}

Note that Case~A has the comparable amounts of $\Msol$ and $\fstars$ as that in \sref{subsec:soliton_stars}.
Since this test yields a different result from the previous test, we suspect it may have been caused by insufficient spatial resolution when the star mass fraction is high in the inner halo.
Hence we repeat this test with $1024^3$ spatial resolution and the same stellar distribution.
Both results are consistent.
We check the system energy which is found conserved up to $1.5\%$ and $0.8\%$ errors for low and high resolution runs, indicating numerical convergence.
The results of both resolution runs are presented in \fref{fig:5.0e09x3}.

The first feature comes to sight is the soliton peak height enhanced by one order of magnitude and the soliton's size shrinks about a half at $\tH$.
The final star peak mass density can reach the original soliton mass density, and overwhelms the dark matter mass density around $\RHL$.   
In fact, the star mass density was originally comparable to the soliton mass density inside the soliton in the first $0.2 \tH$, and the compressed soliton fluctuates in amplitude considerably.  
The stars then get ``heat" up in response to the fluctuating soliton potential and escape from the soliton site.
After this initial transient, the soliton mass is dominant over the stellar mass inside the soliton by a factor of $10$.  
During almost the entire period after then, one finds that the soliton peak height barely changes.
Therefore, contrary to the previous single-soliton case, this case strongly evidences that the soliton evolution is irreversible in the presence of a halo.  

\subsubsection{Case~A-2}

In the initial $0.02\tH$ all cold stars rush into the central region inside the soliton where the stellar mass exceeds the soliton mass, and it severely compresses the soliton.  
Is that the reason the soliton acquires a permanent change in view of the soliton's irreversibility?
To examine this issue, we design the second test where the stellar mass is slowly turned on so that when the above situation happens, stars have negligible mass and stars only acquire their full mass long after.
The instant of star collapse occurs at $0.02-0.03\tH$ and we thus linearly turn on the stellar mass in a duration $0.3 \tH$.  
The result shows that the soliton still gets compressed to the same degree as when stars acquiring full mass instantaneously at $t=0$.
After stars acquiring full mass, the subsequent behavior is similar to that in Case~A-1.
We examine the ratio of stellar mass within $\rsol$, $\Mstars(<\rsol)$, and soliton mass $\Msol$ at certain time steps, as shown in \tref{tab:fstars}.
This test thus shows that the degree of final compression in the soliton does not depend on the large impulsive stress from stars, but on the long term stress from stars in some neighborhood of the soliton.

\begin{table}
\caption{Stars fraction within the soliton: $\Mstars(<\rsol)/\Msol$ }
\label{tab:fstars}
\begin{center}
\begin{tabular}{lcccc}
Case & $0.02\tH$ & $0.2\tH$ & $0.4\tH$ & $0.6\tH$ \\
\hline
A-1& 1.18 & 0.52 & 0.37 & 0.31 \\
A-2& 0.13 & 0.60 & 0.47 & 0.29 \\
\end{tabular}
\end{center}
\end{table}

\subsubsection{Case~A-3}

This leads to a third test where we replace the live stars with an fixed external potential, in this case a 3D potential of a de Vaucouleurs surface density. 
The mass fraction of the external potential is the same as the live star case, and the de Vaucouleurs effective radius is $2.2$ times the soliton width, also consistent with the initial stellar configuration of Case~A-1.
We turn on the external potential within a period $0.05 \tH$ and turn off in another $0.05 \tH$.
Like the previous single soliton case, the soliton gets severely compressed when the external potential is on, but recovers its original configuration when the external potential is off, indicative of reversibility.
With this result, we conclude that the instantaneous soliton compression bears little relation to the soliton irreversibility.
The evidence points to that the irreversibility be originated from a long term secular change.

\subsubsection{Long term behavior of $\Msol\cdot\rsol$}

This long-term trend can be best examined by a scale-invariant product, $\Msol\cdot\rsol$, and we look for any signature of long term secular change.  
For convenience, we define a ratio
\be
\label{eqn:r_mc}
R(t)= [\Msol(t) \cdot \rsol(t)]/[\Msol(0) \cdot \rsol(0)],
\ee
and monitor this ratio throughout the evolution.

\begin{figure*}
\includegraphics[width=0.7 \textwidth, angle=-90]{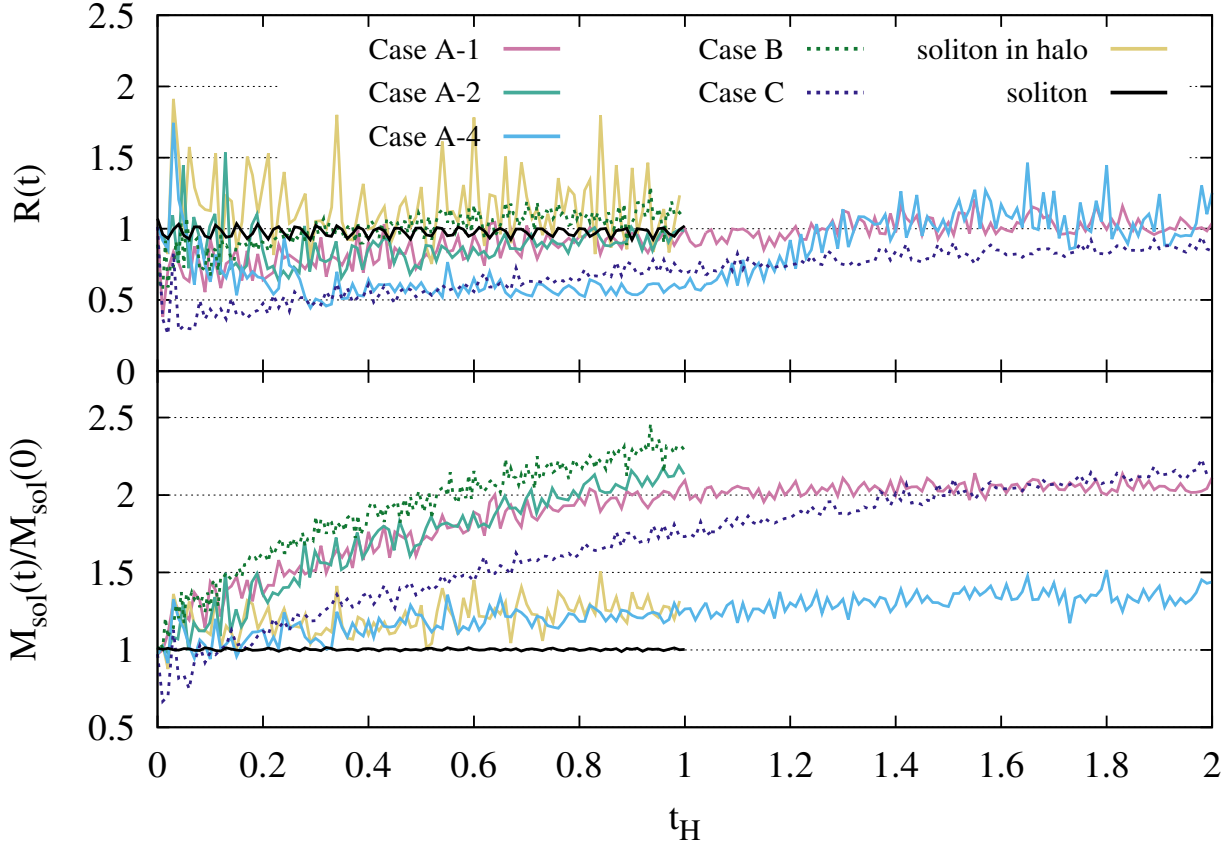}
\caption{
Time evolution of $R(t)= [\Msol(t) \cdot \rsol(t)]/[\Msol(0) \cdot \rsol(0)]$ and $\Msol(t)/\Msol(0)$.
The soliton radius $\rsol$ is defined as the radius where the density is $1/20$ of the peak density and the soliton mass $\Msol\equiv M(<\rsol)$.
Plotted in black and yellow are the references of a single soliton and a soliton in a halo, both having no stars.
Plotted in red, green, cyan, dark green (dot), and dark blue (dot) are results of Cases~A-1, A-2, A-4, B, and C, respectively.
Here one sees Cases~A-1 and A-2 have soliton mass increased by almost a factor $2$, and $R$ resumes unity.
One also finds Case~A-4 is almost reversible where the soliton mass changes little.
Results of Case~B are similar to Case~A-1.
Case~C has a star cluster that inhibits the soliton's ability to absorb mass with the halo, and only after the star cluster is substantially weakened at a later time can the soliton behave normally.
The large fluctuations of $R$ are caused by the less prominent solitons which can be contaminated by their halos during the soliton oscillation.
}
\label{fig:mr_test}
\end{figure*}

For Case~A-1, the soliton initially shrinks in size and the density increases accordingly due to compression.
The scaling relation is violated and $R(0.02\tH) \sim 0.6$.
We unexpectedly find that the soliton can in the long run manage to restore the scaling relation by gaining mass from the halo.
On the way of gaining mass, the soliton also clears stars inside it.
The soliton however does so slowly, and by the time it almost reaches a steady state at $\tH$, the ratio $R(\tH)=0.93$ and the stellar mass within the soliton is almost one order of magnitude below the soliton mass.
From $\tH$ to $2\tH$, the stellar mass is turned off to be zero and the halo is relaxed to another steady state.
But we find the state does not change during this period,  
showing $R(\tH\to2\tH)$ close to $1$.
The soliton mass remains the same during this period as well.
The result is plotted in red curves of \fref{fig:mr_test}.
We check $R(t)$ for Case~A-2 as well, and once stars acquire full mass, the trajectory of $R(t)$ is identical to Case~A-1.
The result is plotted in cyan.
For references, we also plot $R(t)$ for a single soliton (black curve) and for the soliton with halo (yellow curve), both having no stars.
It is noted that the large fluctuations of $R$ are caused by the less prominent solitons which can be contaminated by their halos during the soliton oscillation.
This observation provides a deeper understanding of the long-term irreversibility: after the soliton gains weight, it would not give up the weight despite the stress exerted by stars are released.

\subsubsection{Case~A-4}

Finally, we test a case with the aforementioned de Vaucouleurs potential replacing live stars for a long term evolution.
We do the same as the live star case, turning on the potential gradually from $0$ to $0.3\tH$, letting the system relax till $\tH$, then turning off the potential gradually after in another $0.3 \tH$ and letting the system relax till $2\tH$.
Plotted in cyan curves of \fref{fig:mr_test} is its result.  
The ratio $R<1$ when the external potential is applied, and $R=1$ when the external potential is removed.  
Time reversibility is recovered!
Perhaps more surprisingly, we find the soliton mass change little throughout the evolution.
That is, the soliton cannot exchange mass with the halo when stars inside the soliton are not allowed to coevolve.  
This case is in contrast to tests where live stars inside the soliton can irreversibly get heated up and escape.

\subsubsection{Granule sizes}

\begin{figure*}
\includegraphics[width=0.7\textwidth, angle=-90]{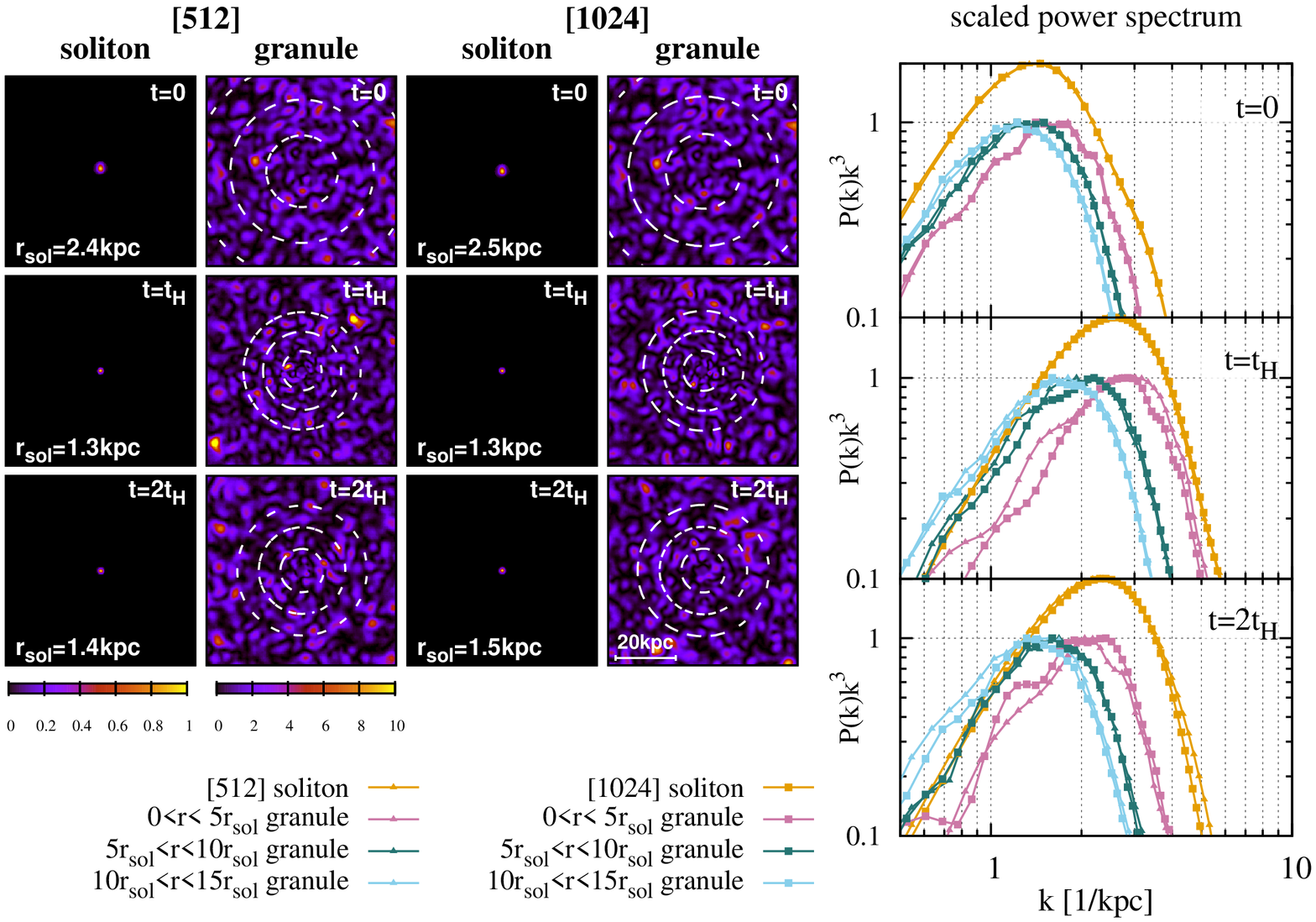}
\caption{
The size of solitons/granules of Case~A-1.
The triangle and square curves are from the $512^3$ and $1024^3$ simulations respectively.
The left four columns are the slices of solitons/granules at different evolution stages with the $512^3$ and $1024^3$ resolutions.
The color boxes show the relative intensity of soliton to the $\rho_{\rm peak}$ (left) and the relative intensities of granules to the local density profile (right).
Both color boxes are in linear scale.
The right column shows the arbitrarily scaled power spectra of
(1) a solitonic core (orange),
(2) granules within $0<r<5\rsol$ (red),
(3) granules within $5\rsol<r<10\rsol$ (green), and (4) granules within $10\rsol<r<15\rsol$ (cyan) evaluated at $t=0$ (top-right), $t=\tH$ (middle-right), and $t=2\tH$ (bottom-right).
Each region of granules is in between dash circles on the left.
The triangle/square curves show the results of $512^3/1024^3$ resolutions on the right.
The peak of power spectrum is able to characterize the size of density fluctuation in a field.
The solitonic core becomes narrower by about a factor of $2$ from $t=0$ to $t=\tH$.
The granules at various radii also become narrower as shown by the right-ward shifts of all spectra.
When stars become massless at $t=2\tH$, the spectral peaks of granules shift back toward the left but the soliton remains little affected.
The granule sizes in different radii reveal how non-isothermality distributes in the inner halo.
}
\label{fig:5.0e09_ps}
\end{figure*}

These all happen in the presence of a halo.
Does it means a secular change also occurs in the halo, particularly to the halo granules?
To address this question, we inspect the power spectra of granules of Case~A-1 at various halo radii as shown in \fref{fig:5.0e09_ps} with $512^3$ and $1024^3$ resolutions.
To clearly see the density fluctuation of granules, we first normalize the density field to the average halo profile $\rho(r)/\langle\rho\rangle(r)$, which is the relative intensities of granules to the local density.
The normalization basically makes the soliton close to unity.
We also excise the solitonic core within $\rsol$ to isolate the soliton. 
In the left four columns, we present the slices of solitons/granules at $t=0$ (top), $t=\tH$ (middle), and $t=2\tH$ (bottom).
The color boxes show the relative intensity of soliton to the $\rho_{\rm peak}$ (left) and the relative intensities of granules to the local density (right).
We then subtract off unity from the granular density to be $\delta\rho/\langle\rho\rangle$ and segregate the density fluctuation within three spherical shells:
(1) $0<r<5\rsol$ (red), (2) $5\rsol<r<10\rsol$ (green), and (3) $10\rsol<r<15\rsol$ (cyan).
The ranges of shells are bound by dashed circles in the sliced images of granules.
In other words, each time when we calculate a power spectrum, the density fluctuation in a given region remains and others are set to be zero.
The power spectra at different evolution time are shown in the right column.
For reference we also show the power spectra of the soliton in orange. 
(The vertical axis is scaled to arbitrary values to help visualization of peak positions.)
If the halo should assume a different mass $M_{\rm hX}$, the horizontal axis, the wave number, is changed by a factor $(M_{\rm hX}/M_{\rm h0})^{1/3}$.
The triangle/square spectra show results from 512/1024 resolutions, which track each other quite well demonstrating numerical convergence.
The peak position of power spectrum can characterize the size of density granules in the halo.

One observes that as time progresses, not only the central soliton but also granules become narrower as all spectral peaks shift toward high $k$. 
Moreover, granules become ``non-isothermal" after evolving from the initial almost ``isothermal" distribution.
When stars become massless at $t=2\tH$, the spectra of granules shift back to the low $k$ to various degrees, but the soliton has the smallest shift.
Despite all the changes, we however notice that the spectral peak of the inner most shell is always close to the soliton spectral peak, indicative of the soliton and the inner most halo sharing the same effective ``temperature".

\subsubsection{Cases~B and C}

\begin{figure*}
\includegraphics[width=0.7 \textwidth, angle=-90]{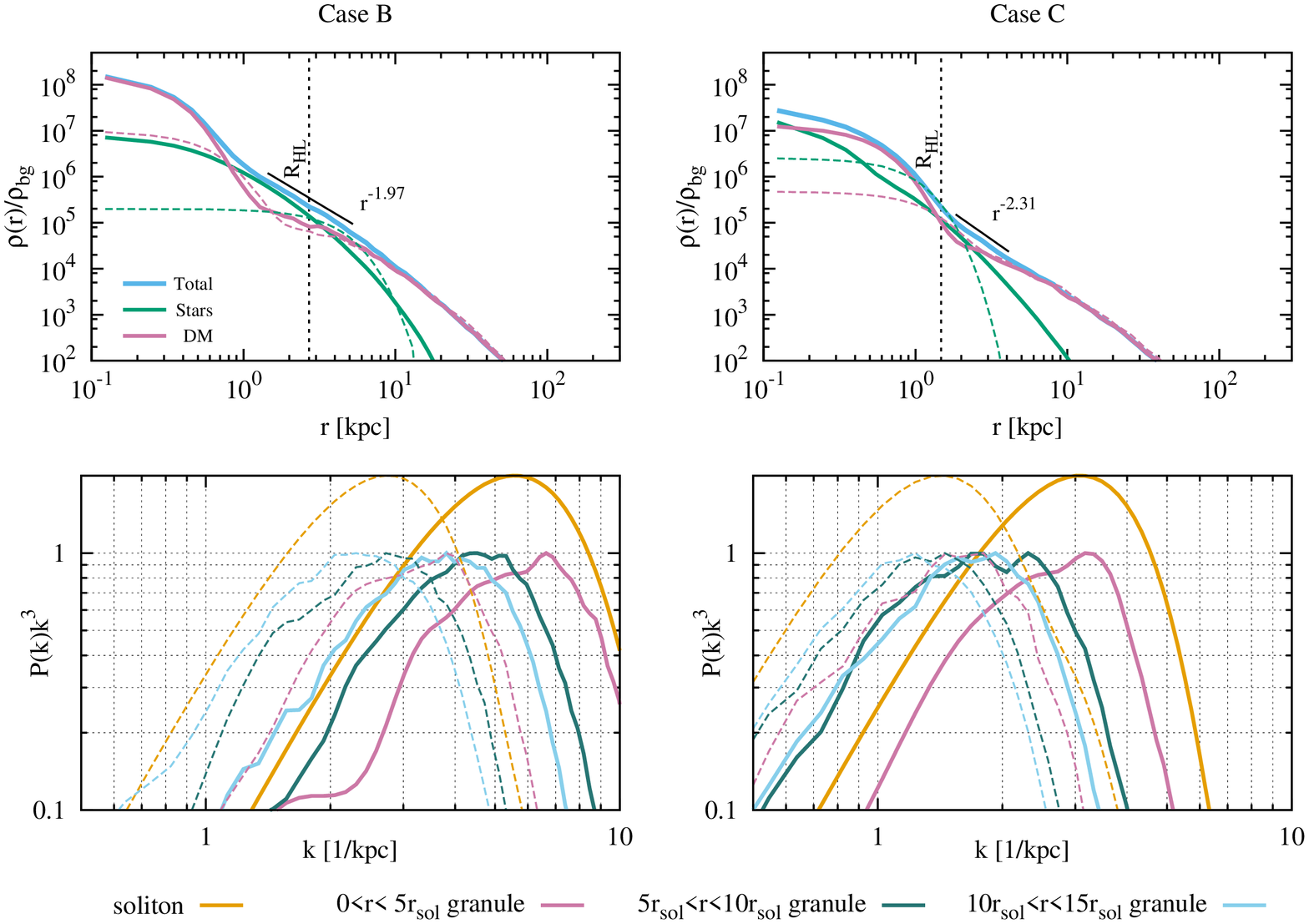}
\caption{
Results of Cases~B and C at $\tH$.
Density profiles are shown in the top panels, and arbitrarily scaled power spectra are shown in the bottom panels.  
The dashed curves of power spectra show the initial state ($t=0$) of different annuli, and the solid line at $t=\tH$, which are similar to Case~A-1. 
Also similar to Case~A-1, the velocity dispersion of Case~B in the insert rapidly increases inward by a factor $2$ from $\RHL$ caused by the soliton potential.
The difference in velocity dispersion profiles of Cases~B and C at the very center is due to the existence of a cold, self-bound star clusters inside the soliton in Case~C.
}
\label{fig:others}
\end{figure*}

We proceed to examine Cases~B and C, where we turn on stars immediately at $t=0$ and let the star-halo system to relax till $\tH$ as similar to  Case~A-1, as shown in \fsref{fig:others} and dark green and dark blue (dot) curves of \fsref{fig:mr_test}. 
Case~B has fairly consistent result as Case~A-1, but increases more soliton mass due to higher $\fstars$.
The soliton gains mass from halo slowly as it removes stars within.  
The final soliton profile obeys the soliton scaling, and its final soliton mass is a factor of 2 of the initial mass and the density increases by one order of magnitude.  
The halo granules also behave the same as those in Case~A-1.

The initial core radius of the bare soliton is $\rcore=0.65$~kpc for Case~B, the half-light radius of the final relaxed state is $\RHL=2.7$~kpc, and hence $\RHL/\rcore\simeq 4$.  
Scaled up to a halo of $1.2\times 10^{11}\Msun$, we have the bare soliton mass $\Msol\simeq 10^{9}\Msun$ and $\Mstars = 6.0\times 10^9\Msun$, consistent with \eref{eqn:ltomh}.  
The bare soliton core radius of a halo of this mass is $0.33$~kpc and hence with $\RHL/\rcore=4$, we have $\RHL\simeq 1.3$~kpc, also consistent with the observed half-light radius of $1.4\times10^{11}\Msun$ galaxies.

With a close examination of Case~C, we however find the dynamical result inconsistent with previous cases, in that the ratio $R(\tH)$ is substantially smaller than unity by a non-negligible margin, as shown in the dark blue (dot) curves of \fref{fig:mr_test}.
Since stars are initially all inside the soliton, the origin of this inconsistency turns out to be related to a self-bound star cluster residing within the soliton, and hence the soliton is incapable to remove during the available time span.
Like the external potential test, the presence of a star cluster inhibits the soliton to exchange mass with the halo.  
To demonstrate the effect really arises from the star cluster, we let Case~C continue to run till $2\tH$.  
By then the star cluster becomes weakened and the soliton finally resumes what it ought to be.
We will return to the issue of the star cluster in the next section.   

Using the halo scaling for Case~C, this galaxy is less massive than the scaled halo of Case~A-1 since $\fstars$ is only a half. 
\eref{eqn:ltomh} for the star fraction demands that the mass of the scaled halo is about $30\%$ less than that of Case~A-1, i.e., $5.0\times 10^{10}\Msun$.  
The rescaled half-light radius is also smaller, $\sim 0.8$~kpc at $\tH$, representing a peculiar situation where star distribution are so compact that a self-bound star cluster can form in the inner halo.   

\subsubsection{Summary}

\begin{table*}
\caption{Simulation setups}
\label{tab:tests}
\begin{center}
\begin{tabular}{lllccc}
Case & halo & initial stellar distribution & $\fstars$ & turn-on $\Mstars$& turn-off $\Mstars$ \\
\hline
A-1 & \psidm\ $5.0\times10^{ 9}\Msun$  & $3$ times wider than the soliton core               & 3   & $t=0$         & $\tH - 1.3\tH$\\
A-2 & \psidm\ $5.0\times10^{ 9}\Msun$  & same as Case A-1                                    & 3   & $0 - 0.3\tH$  & $\tH - 1.3\tH$\\
A-3 & \psidm\ $5.0\times10^{ 9}\Msun$  & fixed de Vaucouleurs potential ($\RHL=5.5$~kpc)     & 3   & $0 - 0.05\tH$ & $0.05\tH - 0.1\tH$\\
A-4 & \psidm\ $5.0\times10^{ 9}\Msun$  & same as Case A-3                                    & 3   & $0 - 0.3\tH$  & $\tH - 1.3\tH$\\
B   & \psidm\ $1.7\times10^{10}\Msun$  & $6$ times wider than the soliton core               & 6   & $t=0$         & -             \\
C   & \psidm\ $5.0\times10^{ 9}\Msun$  & half of the soliton core                            & 1.5 & $t=0$         & $\tH - 1.3\tH$\\
A$'$&    CDM\ $5.0\times10^{ 9}\Msun$  & same as Case A-1                                    & 3   & $t=0$         & -             \\
B$'$&    CDM\ $1.7\times10^{10}\Msun$  & same as Case B                                      & 6   & $t=0$         & -             \\
\end{tabular}
\end{center}
\end{table*}

We sum up all above tests for the dark matter halo with stars in the inner halo in \tref{tab:tests}.
The results illustrate several important points about the dynamics of the soliton$+$halo$+$star system.
\begin{enumerate}[(i)]
\item Unlike the previous single soliton test, the soliton manages to preserve the scaling relation, $\rho_{\rm peak}\propto\rcore^{-4}$,
in the presence of a halo and live stars.  
But it takes a long time for the soliton to relax, to exchange mass with the halo and also to clear the stars inside to recover this scaling relation.
\item Also unlike the previous single soliton case, the soliton exhibits irreversibility, and the system presents a hysteresis.
\item The halo becomes ``non-isothermal" upon adding stars with the inner halo hotter than the outer halo.
After the removal of stars, the non-isothermality persists.
In the live star case, we further find that the final soliton$+$halo system at $2\tH$ has a more negative energy than the initial soliton$+$halo system before adding stars by a factor $1.6$.
The dark matter halo is losing energy to the kinetic energy of removed hot stars.
\item When the added stars are not live and do not coevolve with the soliton$+$halo system, the soliton mass does not change and the soliton evolution is time reversible, like the single soliton case.
The role of a strong bound star cluster inside the soliton can be similar to an external potential in inhibiting the soliton to relax.
\item From (iii) and (iv), it is clear that stars in the presence of a halo play a vital role in the irreversibility of the soliton.
Stars are irreversibly heated by the time-dependent halo$+$soliton potential, in such a way that stars at $\tH$ have more positive energy than stars initially by absorbing the energy of dark matter halo.
\end{enumerate}

\section{Stellar Kinematics}
\label{sec:stellar_kinematics}

\subsection{Logarithmic slope $r^{-2}$ and projected surface density}
\label{subsec:slope_sersic}
In this work, we adopt the half-light radius $\RHL$ as the yardstick to measure the local logarithmic slope.
We apply least-squares fitting to the total mass profile in $0.5 \lesssim r \lesssim 2.0 \RHL$ but exclude the data in $r < \rsol$.
In \fsref{fig:5.0e09x3} and \ref{fig:others}, we find the local logarithmic slopes of the total mass density at $\RHL$ are close to $-2$ to various degrees, and more so when $\RHL$ is farther away from the soliton as it is not contaminated by the tail of the soliton.
Also near $\RHL$, the stellar density is locally comparable to, or even slightly greater than, the dark matter density, reflecting that this particular value of local slope is equally contributed by stars and dark matter.

In our tests of other parameters with wider initial stellar distributions reveal that most halos also exhibit $r^{-2}$ local slopes near $\RHL$.  
The cases that fail to exhibit this feature are either the initial star distribution are too compact, such as Case~C, or too spread out.  
The latter is sensitive to the the mass ratio $\fstars$ but the former is not.
When this mass ratio is too low where the stellar density can never be comparable to the dark matter density at $\RHL$, the soliton can barely be compressed and the total mass surely follows the dark matter distribution, thus the slope becoming flatter than $-2$.  
We summarize all test results in \aref{app:slope}. 

\begin{figure*}
\includegraphics[width=1.0 \textwidth, angle=0]{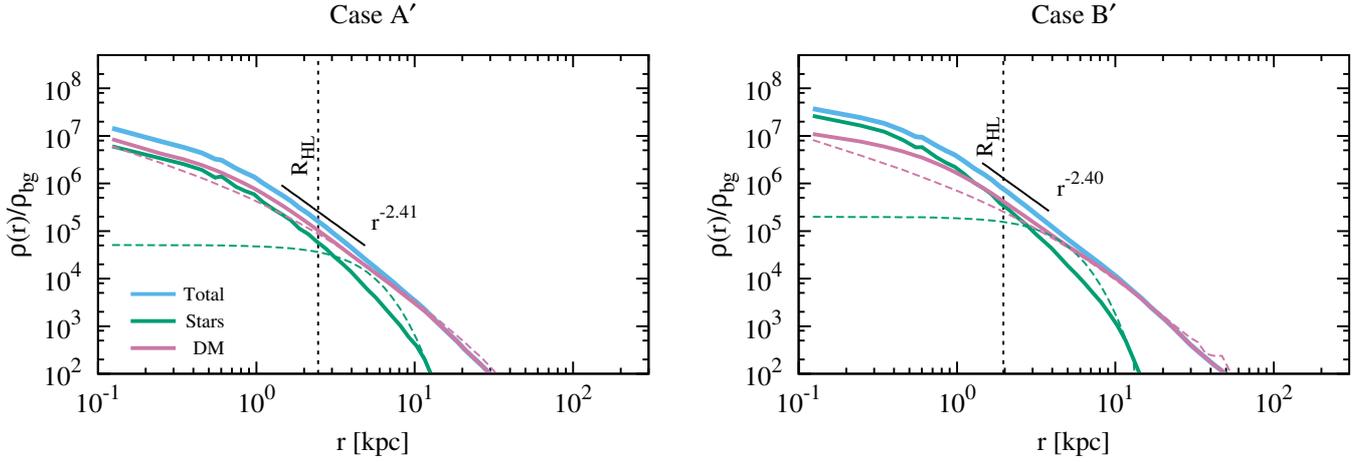}
\caption{
Results of Cases~A$'$ and B$'$ at $\tH$.   
Different from \psidm\ case, stars either overwhelm, or are comparable to, the dark matter at the halo center.  
At $\RHL$, the log-slopes of total star density are about $\sim-2.4$, substantially steeper than $-2$.
}
\label{fig:othersCDM}
\end{figure*}

As mentioned in \sref{sec:intro}, the previous works found the logarithmic slope of inner CDM halo substantially steeper than $-2$.   
To have a fair comparison, here we test CDM halos of the same halo mass, stellar mass and initial stellar distribution as the corresponding \psidm\ halos for Cases~A-1 and B, called Case~A$'$ and~B$'$.   
The CDM results at $\tH$ are shown in \fref{fig:othersCDM}.  
The values of logarithmic slopes at $\RHL$ at $\tH$ are indeed approximately $-2.4$ for both halos, consistent with previous studies for more massive halos \citep{DuttonEtal15}.   

Aside from the logarithmic slope, we find stars in CDM halos tend to either dominate or be comparable to dark matter and are more dense than those in \psidm\ halos at the central region.   
This is in part due to the removal of stars by the soliton, but also in part due to that the granularity of \psidm\ halos also leads to star diffusion in the long run.

\begin{figure}
\includegraphics[width=0.35 \textwidth, angle=-90]{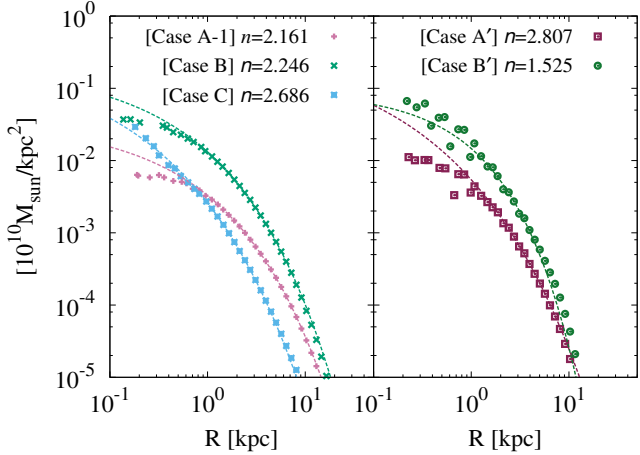}
\caption{
The projected stellar surface densities of Cases~A-1, B, C (left), A$'$ and B$'$ (right).
We fit them by Sersic profiles and the Sersic indices ($n$) are labeled at the top.
}
\label{fig:sersic}
\end{figure}

The star surface brightness is observable.
In \fref{fig:sersic}, we show 2D projected stellar surface densities of Cases~A-1, B, and C discussed above.
We fit them by Sersic profile, and except for the core region the fit is quite satisfactory, with the Sersic index $3>n>2$.  
This range of $n$ is consistent with dwarf ellipticals \citep{Fisher&Drory08,Fisher&Drory10}. 
In the core region where the soliton is located, stars are depleted by the soliton and therefore it presents a deficit from the Sersic fit.
However, Case~C is an exception that shows no deficit.
As we noted earlier, Case~C has a self-bound star cluster inside the soliton and this star cluster is hard for the soliton to remove in a finite time.  
The Sersic indices of CDM halos are also shown for comparison.  

According to our halo scaling, the parameters of Case~A-1 represent a $7.0\times10^{10}\Msun$ halo, Case~B is a substantially more massive halo ($1.2\times10^{11}\Msun$), and Case~C the least massive halo  ($5.0\times10^{10}\Msun$).
But Case~C has the highest value of $n$ among the three, a situation that is opposite to the observational trend of Sersic index where $n$ increases with stellar mass \citep{Fisher&Drory08,Fisher&Drory10}. 
Perhaps a sample with only three numbers is not representative, but it may also be suggestive of some additional physics missing in our tests for lower mass halos where stellar mass is observed to decrease quite rapidly when halos are below $10^{11}\Msun$\citep{Vale&Ostriker04}.
One example of the missing physics in our tests is tidal stripping by environments.

\subsection{Velocity dispersion and star cluster}
\label{subsec:star_cluster}

\begin{figure}
\includegraphics[width=0.35 \textwidth, angle=-90]{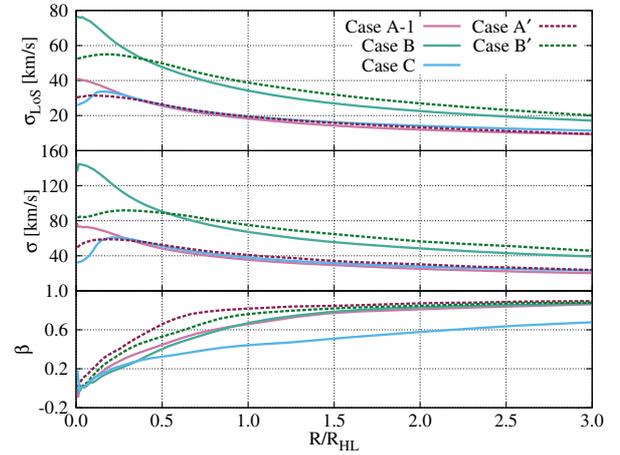}
\caption{
The line-of-sight velocity dispersion profiles $\sigma_{\rm LoS}$ (top), the 3D velocity dispersion profile $\sigma$ (middle), and the velocity anisotropy profile $\beta$ (bottom) of Cases~A-1, A$'$, B, B$'$, and C with distances scaled to $\RHL$.
}
\label{fig:vel}
\end{figure}

\fref{fig:vel} summarizes what we will discuss in this sub-section.
We find the stellar 3D velocity dispersion $\sigma$ and 2D projected (line-of-sight) velocity dispersion $\sigma_{\rm LoS}$ can reach roughly peak velocity dispersions $75$~km/s and $42$~km/s, respectively, for Case~A-1 of a halo $5.0\times10^9\Msun$, and roughly peak velocity dispersions $130$~km/s and $70$~km/s, respectively, for Case~B of a halo $1.7\times 10^{10}\Msun$.
When we scale the halo as having a mass $10^{11}\Msun$, the velocities become a factor $(M_{\rm hX}/M_{\rm h0})^{1/3}=2.7$ greater for Case A-1 to reach $200$~km/s and $115$~km/s, respectively, and a factor 1.8 greater for Case~B reaching $235$~km/s and $125$~km/s, respectively.
These values are reasonable for galaxies of stellar mass several times of $10^9\Msun$ extrapolated from more massive elliptical galaxies with $\sim10^{10}\Lsun$, assuming that the stellar mass-to-light ratio is $3$ times the solar value \citep{FranxEtal89}.

The abrupt increase of velocity dispersion toward the galactic center inside $\RHL$ by a factor of $2$ can be the key feature of evidence for the soliton as shown in \psidm\ halos A and B.  
Here the gravity of the central massive object is at work for acceleration.
The velocity dispersions for CDM halos A$'$ and B$'$ also exhibit gentle inward increase, but they are flattened out at the very center due to the lack of distinct central massive lump.   
High angular resolution spectroscopic observations ought to tell the difference.

\begin{figure*}
\includegraphics[width=1.0 \textwidth, angle=0]{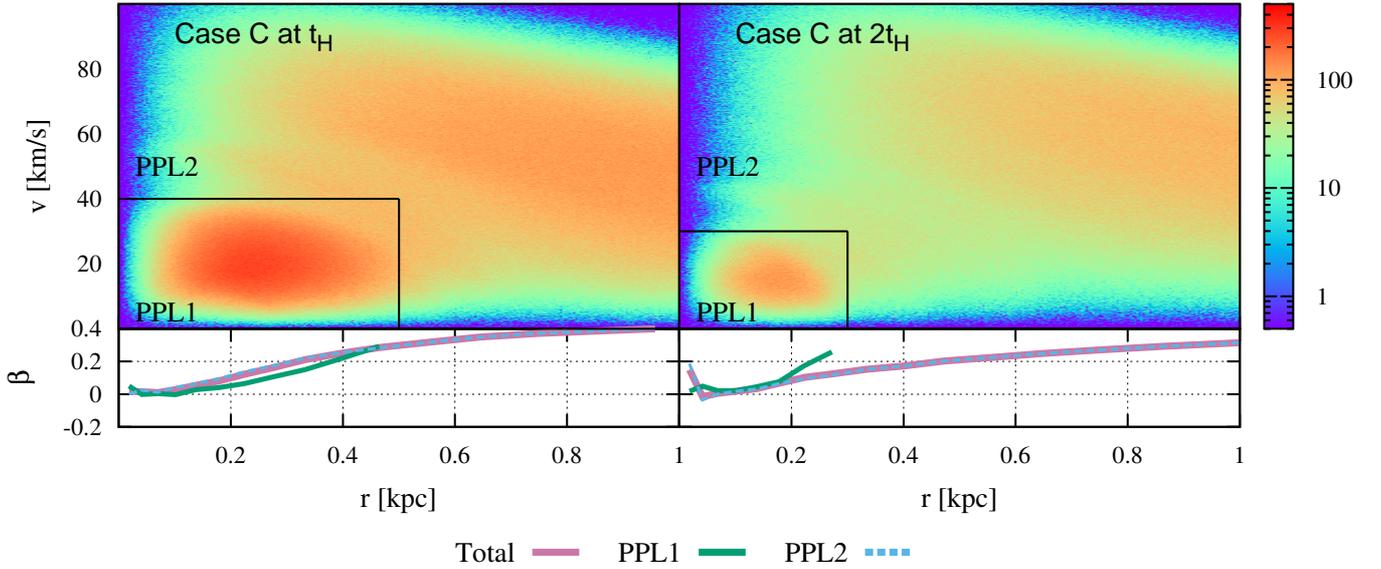}
\caption{
Stellar phase-space distributions of Case~C at $\tH$ and $2\tH$.
Two populations of stars co-exist: a cold, self-bound star cluster located inside the soliton (PPL1), and stars in halo with high velocity dispersion (PPL2).
At $2\tH$, the right-bottom sector of PPL1 clearly shows weakening of the star cluster; these stars are not sufficiently abundant to be self-bound, and they are bound by the soliton potential.
}
\label{fig:phase_space}
\end{figure*}

But for Case~C, as shown in the cyan curve, where a self-bound star cluster exists inside the soliton, the velocity dispersion has a dip in the central region coinciding with the soliton location.
This dip is primarily contributed by the star cluster with much lower velocity dispersion.   
To see the star cluster, we examine the phase space $(|{\bm v}|, r)$ distribution of stars.
Shown in \fref{fig:phase_space} is a phase-space plot for stars of Case~C.  
The top left panel is the distribution at $\tH$ and top right panel at $2\tH$.  
It clearly reveals phase separation of two populations on both panels.
At $\tH$, the star cluster, which contains 10\% of total stars, residing inside the soliton has very low velocity dispersion about $20$~km/s as opposed to $\sim70$~km/s for halo stars.
At a later time $t=2\tH$, the star cluster is substantially weakened due to the heating of the unstable soliton fluctuating potential.  
The remaining stars in the star cluster, which are now only 2\% of the total, have an even lower velocity dispersion about $10$~km/s.
This low-density star cluster at $2\tH$ is confined by the soliton potential, as the star self-gravitating is unable to bound the stars.  
We find the ratio of the star kinetic energy to potential energy approximately equals $-1$ for the weak star cluster at $2\tH$, while for the strong star cluster at $\tH$, it is self-bound and has a virial ratio approximately equal to $-0.5$.

The bottom panel of \fref{fig:vel} plots the velocity anisotropy $\beta$ of stars.
Not unexpectedly, the star velocity anisotropy of halo stars is normal, radial in the outer halo and isotropic in the inner halo.  
The weak star cluster at $2\tH$ in the bottom right panel of \fref{fig:phase_space} is slightly more radial with anisotropy reaching $\beta \sim 0.3$ at the outer edge of the cluster.
This may be an indication of stripping of loosely bound stars at the outer edge.  
But these stars are contaminated by halo stars, and hence only an unquantifiable statement can be made.

\section{Conclusion}
\label{sec:conclusion}
In this work we consider a simple \psidm\ halo model incorporating stars in the inner halo and explore the dynamical interplays between the two species.
Some fundamental issues such as reversibility/irreversibility of the soliton in response to environmental changes are investigated via a series of controlled simulations using dwarf ellipticals with Sersic index $n$ between $2$ and $3$ as a test bed.
We find that the presence of halo and stars plays an essential role for the soliton dynamics.
Soliton absorbs mass from the halo to satisfy the scaling relation when stars can be cleared out of the soliton.
As a result of the increasingly massive soliton, the shape parameter with stars increases about an order of magnitude compared with that without stars.

Soliton actually has an effective mean to clear these stars out of the way by its fluctuating potential.
But once some halo mass gets absorbed, the soliton does not return its original mass after stars are totally eliminated, exhibiting irreversibility as opposed to reversibility of a single soliton with no halo.
Though the way we remove stars is somewhat artificial, the essence of irreversibility should be preserved.
One can imagine a physical situation where most stars are removed by tidal stripping so that the stellar mass fraction in the stripped halo is negligibly small, such as those in dwarf spheroidal galaxies.
Despite the very low stellar mass fraction, such a soliton can no longer be the soliton derived from pure \psidm\ simulations.
The former can be more massive than the latter by, for example, a factor of $2$ due to the irreversibility.
This finding is purely empirical, and the underlying physical interpretations still remain to be investigated.

We also find that the local logarithmic slope of the composite mass density of dark matter and stars
can approach isothermal with power index $-2$ at the 2D half-light radius, provided that the half-light radius is at distance from the soliton, and that the stellar mass in the inner halo is comparable to the dark matter mass.  
This holds true normally for more massive halo where the soliton and the half-light radius are well separated and the star fraction is moderate to high in the inner halo. 
(See \aref{app:slope}.)

Inward sharp rise of the line-of-sight velocity dispersion toward the galactic center has often been observed in elliptical galaxies and in bulges \citep{SagliaEtal10,NessEtal13,ZoccaliEtal14,BoardmanEtal17}, but so far the interpretation involves the presence of a supermassive black hole \citep{Kormendy&Ho01} or self-gravitying stars \citep{Binney80, PortailEtal17}.
In particular, \cite{PortailEtal17} have found that such a sharp peak of velocity dispersion in the central $100$~pc of Milky Way can only be explained by an unseen compact stellar bar/bulge of $2\times10^9\Msun$. 
This work offers an alternative interpretation of these observations.
We find that the velocity dispersion, accelerated by the soliton within the half-light radius, can rapidly increase inward by a factor of $2$, which is consistent with observations \citep{SalinasEtal12}.
For more massive elliptical galaxies, we expect the rise of velocity dispersion is even sharper as the soliton becomes heavier and narrower where the soliton potential is deeper.
It can be a key star kinematic feature in favor of the presence of a soliton.

A cold star cluster is found to be phase-separated from and coexists with halo stars with high velocity dispersion in one of our simulation runs, Case~C, which corresponds to the lowest mass halo with the smallest half-light radius of the halos under study.
In the framework of \psidm\, most stars should be depleted by, for example, tidal stripping in satellite dwarf spheroidal galaxies, and the remaining stars detected are a cold population with low velocity dispersion, as that in Fornax Dwarf, confined inside the soliton potential \citep*{SchiveEtal14}. 
These cold stars resemble the star cluster of Case~C at $2\tH$.
Details about how the tidal stripping, or other mechanisms, depletes stars in satellite dwarf spheroids are still unclear and require further investigations.

\section*{Acknowledgements}

We thank D.~J.~E.~Marsh for carefully reading our manuscript and for giving constructive comments.
This work is supported by the grant, MOST 103-2112-M-002-020-MY3.
J.~H.~H.~Chan thanks S.~H.~Suyu ,A.~Y{\i}ld{\i}r{\i}m, and Y.~Revaz for useful discussion, and the Max Planck Society for support through the Max Planck Research Group of S.~H.~Suyu.
We thank Ui-Han~Zhang for programming support.

\appendix
\section{Inner Halo scaling}
\label{app:scaling}

Throughout this work we have claimed that halos below $10^{11}\Msun$ can follow a scaling relation.
This premise is rested on similar shape parameters, i.e., similar shapes of different inner halos with respect to solitons shown in Figure 3 of SCB14. 
The shape parameter is defined as the ratio of the soliton peak height to the halo peak height, and for most halos of mass in between $5.0\times10^9\Msun$ to $10^{11}\Msun$, the shape parameter is approximately $100-150$. 

\begin{figure*}
\centering
\includegraphics[width=0.7 \textwidth, angle=-90]{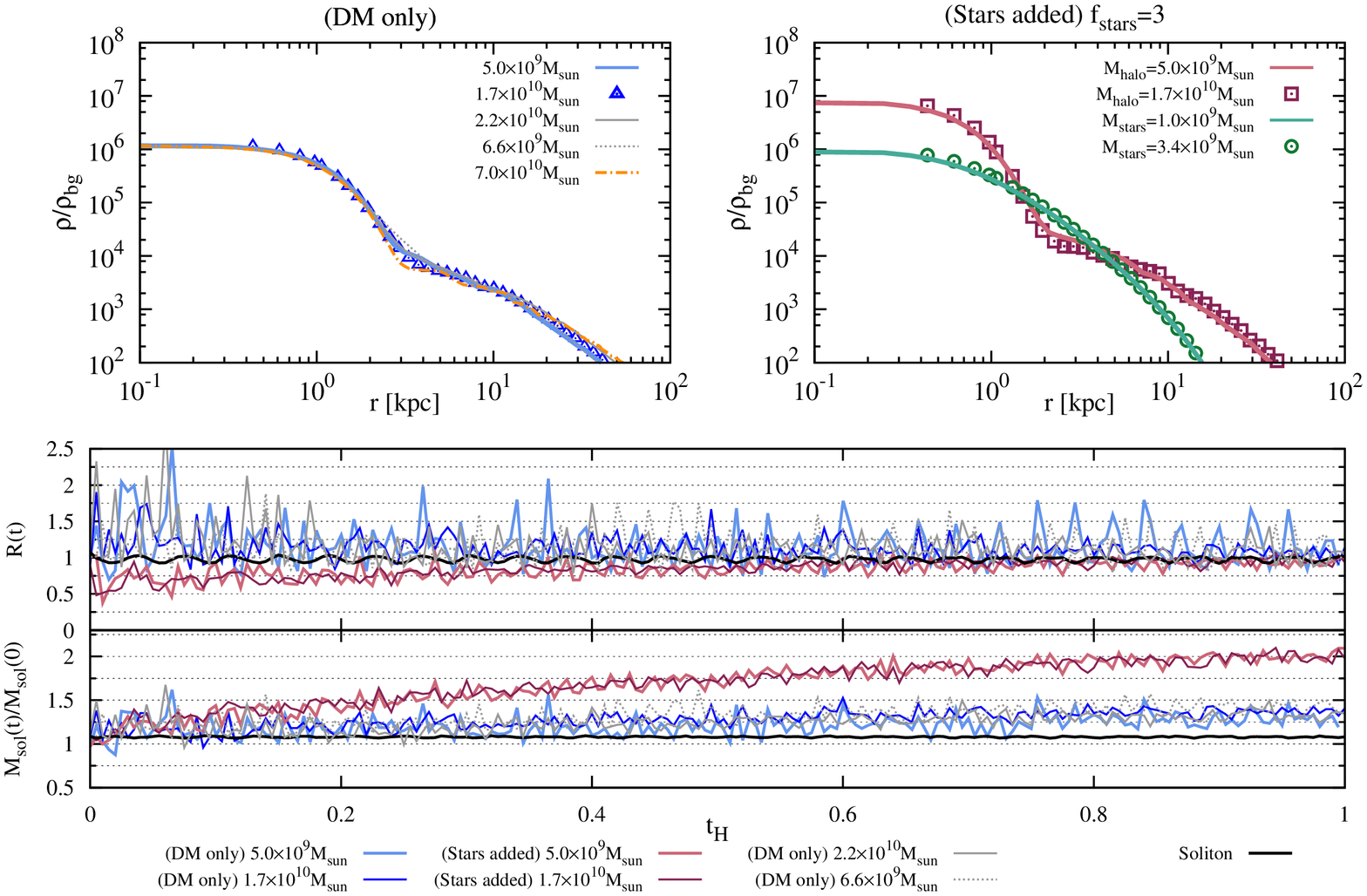}
\caption{
A scaling relation for different mass halos.
The scaled density profile are shown in the upper panel, where $5$ curves are overlapped very well on the left and initially different shape parameters come to convergence after halo relaxation.
The top right shows the scaling resulting of two stars-added halos, and the dark matter and stars also show excellent agreement between the two.  
The evolution of $R(t)$ and scaled $\Msol(t)/\Msol(0)$ of the $5$ halos is shown in the lower panel.
}
\label{fig:scaling}
\end{figure*}

There are however a couple of halos extracted from cosmological simulations that have shape parameters a factor of few below $100$.
The deviations are caused for example by recent mergers, and halos are not sufficiently relaxed.
We therefore extract all $5$ halos from SCB14 in the desired mass range and evolve these halos for another $\tH$ treating them as isolated halos.
After evolution, these halos indeed all relax to shape parameters close to $100$.
On the left panel of \fref{fig:scaling} we show all $5$ halos of masses
$5.0\times 10^9\Msun$, $6.6\times 10^9\Msun$, $1.7\times10^{10}\Msun$, $2.2\times10^{10}\Msun$ and $7.0\times10^{10}\Msun$.
These halo are scaled down to $5.0\times 10^{9}\Msun$ to check the agreement.
The scaling relation of the inner halos follows $r \to \lambda r$ and $\rho \to \lambda^{-4}\rho$ with a free parameter $\lambda$.
The agreement of these halos to the inner halo scaling shown in the top left panel of \fref{fig:scaling} is excellent.
This scaling relation is obeyed by the soliton discussed in \sref{subsec:psidm} and it is of some surprise that inner halo profiles also follow this relation.
In the bottom panels we also show the evolution $R(t)$ and soliton mass ratio $\Msol(t)/\Msol(0)$ of these halos.
They all behave as expected with constant $R$ and $\Msol$.

Why does the shape parameter assume $\sim 100$, and why not $1000$?
Is this halo mass range special, or can the inner scaling be extended to other mass range with a different value of shape parameter?
These are questions awaiting further investigations. 
However, within the mass range in this study, the shape parameter is helpful to constrain the halo structrue of dwarf-galaxy scale, such as Fornax and Sculptor \citep{Mash&Pop15}.

As the current simulation is able to resolve the $5.0\times10^9\Msun$ and $1.7\times10^{10}\Msun$ halos when stars are included, we perform a further test of these two halos aiming to verify the halo scaling even with stars.
We add the same mass ratios of stars relative to the soliton $\fstars=3$ to the $1.7\times10^{10}\Msun$ halo as that in Case~A-1.
Moreover, the initial cold star distribution relative to the initial soliton is chosen the same as Case~A-1, three times wider.   
The simulation run lasts for $\tH$ so that we can compare it with the density profile of \fref{fig:5.0e09x3}.
Plotted on the top right panel of \fref{fig:scaling} are the two dark matter halo profiles and the two stellar profiles.
Here the more massive halo is again scaled down to the less massive halo to check the agreement.
The agreement is also excellent.
We also show the almost identical evolution of $R(t)$ and $\Msol(t)/\Msol(0)$ for the two halos in the lower panel.

The conclusion of this \aref{app:scaling} is that the inner halo of \psidm\ within a certain range of mass ($\sim 5.0\times10^9 \to 10^{11}\Msun$) obeys a universal inner halo profile even when stars are included.

\section{More on logarithmic slope}
\label{app:slope}

In \sref{subsec:slope_sersic}, we have shown the total mass density has a trend to locally follow the $r^{-2}$ isothermal profile at the half-light radius $\RHL$.
The trend becomes clearer when $\RHL$ is larger than the soliton radius $\rsol$ and when the stellar density is comparable to the dark matter density at $\RHL$.
Here we show results of few more simulation runs with different $\fstars$ and different ratios of the initial width of stellar distribution to the initial soliton core width to substantiate our point. 

We choose $\fstars = 3, 1.5$, and $0.75$.  
We also choose initial stellar distributions $\times0.5$, $\times1$, $\times3$, and $\times6$ wider than the solitonic core.
These additional halos are scaled to the mass range $10^{11}-3\times 10^{10}\Msun$ but with half-light radii substantially larger than $1$~kpc, and represent the observed large scatters of half-light radius around the low-mass end of elliptical galaxies.

\begin{figure}
\includegraphics[width=0.5\textwidth, angle=0]{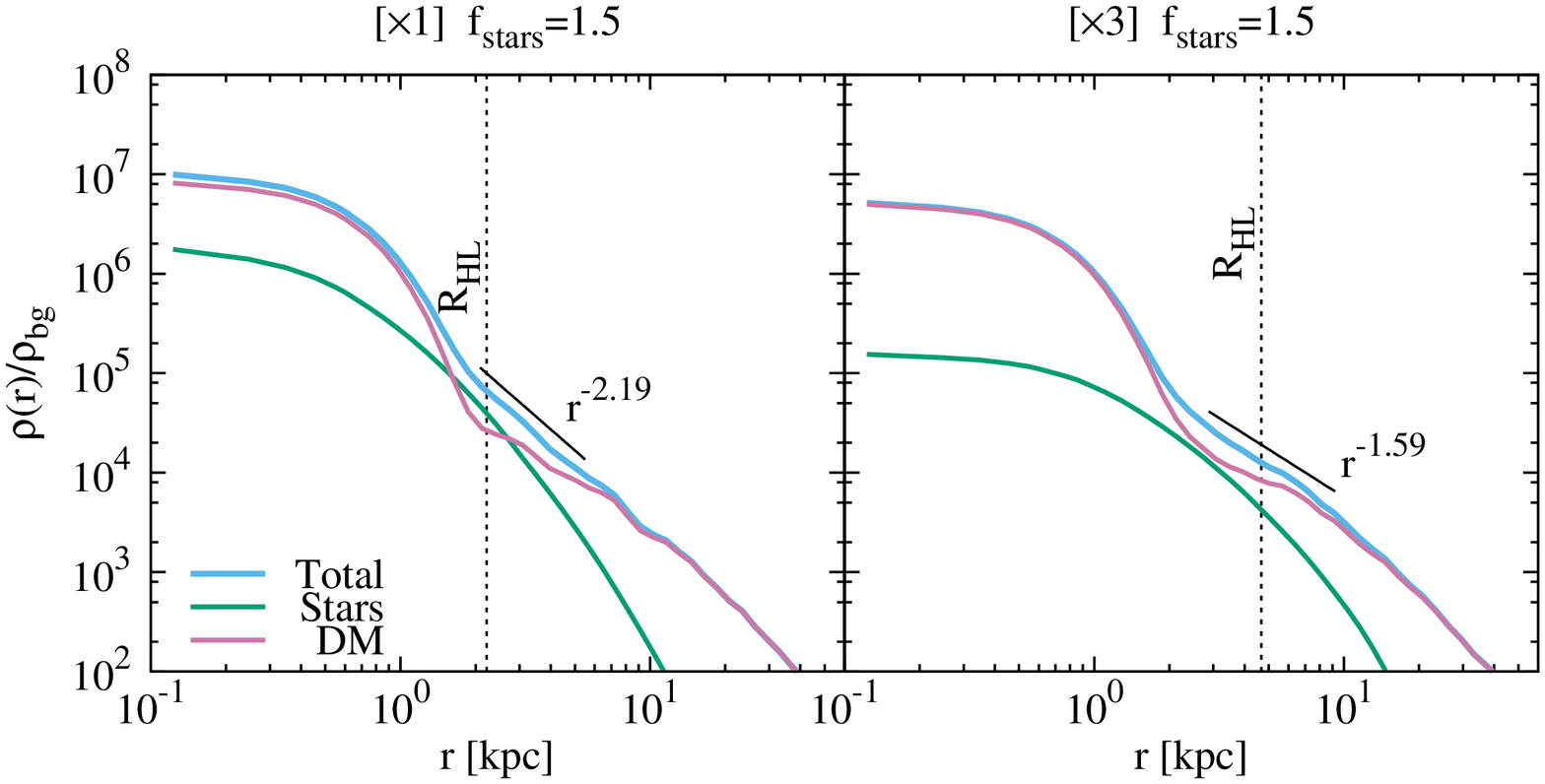}
\caption{
The density profiles succeeding and failing to follow the $r^{-2}$ profile.
}
\label{fig:slope_contrast}
\end{figure}

Listed in \tref{tab:case} is the result of minus logarithmic slopes of all tested halos at $\RHL$, including ones in the main text.
Among the $7$ halos in \tref{tab:case}, one is too flat and the other is too steep compared with others.  
The one halo with too steep a slope has been shown in \fref{fig:others} where $\RHL$ is too close to the soliton.
Here we plot in \fref{fig:slope_contrast} the density profiles of two halos (with asterisks in \tref{tab:case}) as a contrast, the failed one and a success one in following the $r^{-2}$ profile.
The failed has inadequate stars to modify the total mass density in the inner halo.

\begin{table}
\caption{Logarithmic slopes $r^{-\alpha}$ of more tests}
\label{tab:case}
\begin{center}
\begin{tabular}{ccccc}
$\fstars$ & $\times0.5$ & $\times1$ & $\times3$ & $\times6$ \\
\hline
    3 & - & - & 1.95 & 2.01 \\
  1.5 & 2.31 & 2.19$^{*}$ & 1.59$^{*}$ & - \\
 0.75 & 2.05 & 1.84 & - & - \\
\end{tabular}
\end{center}
The halo mass is $5.0\times10^{9}\Msun$. 
The initial stellar distributions are $\times0.5$, $\times1$, $\times3$, and $\times6$ wider than the solitonic core, in company with $\fstars = 3, 1.5$, and $0.75$.
Density profiles of the two halos with asterisks are plotted in \fref{fig:slope_contrast}.
\end{table}

\bibliographystyle{mnras}
\bibliography{psiDMnStars}

\bsp	
\label{lastpage}
\end{document}